\documentclass[journal]{IEEEtran}
\usepackage{cite}
\usepackage{amsmath,amssymb,amsfonts,bm}
\usepackage{amsthm}
\usepackage{dsfont}
\usepackage{algorithm}
\usepackage{algcompatible}
\usepackage{subfigure}
\usepackage{graphicx}
\usepackage{textcomp}
 \usepackage{threeparttable}
\usepackage{xcolor}
\usepackage{bbm}
\usepackage[nolist,withpage]{acronym}
\usepackage{array}
\usepackage{tikz}
\usepackage{pgfplots}
\usepackage{dsfont}
\pgfplotsset{compat=newest}

\definecolor{copperrose}{rgb}{0.6, 0.4, 0.4}
\definecolor{azure}{rgb}{0.0, 0.5, 1.0}
\definecolor{ashgrey}{rgb}{0.7, 0.75, 0.71}
\definecolor{chestnut}{rgb}{0.8, 0.36, 0.36}
\definecolor{airforceblue}{rgb}{0.36, 0.54, 0.66}
\definecolor{cadmiumorange}{rgb}{0.93, 0.53, 0.18}
\definecolor{bleudefrance}{rgb}{0.19, 0.55, 0.91}
\definecolor{carolinablue}{rgb}{0.6, 0.73, 0.89}
\definecolor{blue(ncs)}{rgb}{0.0, 0.53, 0.74}
\definecolor{dodgerblue}{rgb}{0.12, 0.56, 1.0}
\definecolor{cssgreen}{rgb}{0.0, 0.5, 0.0}
\definecolor{cadmiumgreen}{rgb}{0.0, 0.42, 0.24}
\definecolor{cadmiumorange}{rgb}{0.93, 0.53, 0.18}
\definecolor{amaranth}{rgb}{0.9, 0.17, 0.31}
\definecolor{bluegray}{rgb}{0.4, 0.6, 0.8}
\definecolor{cerulean}{rgb}{0.0, 0.48, 0.65}
\definecolor{ceil}{rgb}{0.57, 0.63, 0.81}

\usepackage{etoolbox}
\makeatletter
\newif\if@in@acrolist
\AtBeginEnvironment{acronym}{\@in@acrolisttrue}
\newrobustcmd{\LU}[2]{\if@in@acrolist#1\else#2\fi}

\newcommand{\ACF}[1]{{\@in@acrolisttrue\acf{#1}}}

\begin{document}


\begin{acronym}[LTE-Advanced]
  \acro{2G}{Second Generation}
  \acro{3-DAP}{3-Dimensional Assignment Problem}
  \acro{3G}{3$^\text{rd}$~Generation}
  \acro{3GPP}{3$^\text{rd}$~Generation Partnership Project}
  \acro{4G}{4$^\text{th}$~Generation}
  \acro{5G}{5$^\text{th}$~Generation}
  \acro{AA}{Antenna Array}
  \acro{AC}{Admission Control}
  \acro{AD}{Attack-Decay}
  \acro{ADC}{analog-to-digital converter}
  \acro{ADMM}{alternating direction method of multipliers}
  \acro{ADSL}{Asymmetric Digital Subscriber Line}
  \acro{AHW}{Alternate Hop-and-Wait}
  \acro{AI}{Artificial Intelligence}
  \acro{AirComp}{over-the-air computation}
  \acro{AirFL}{over-the-air federated learning}
  \acro{AMC}{Adaptive Modulation and Coding}
  \acro{ANN}{artificial neural network}
  \acro{AP}{\LU{A}{a}ccess \LU{P}{p}oint}
  \acro{APA}{Adaptive Power Allocation}
  \acro{ARMA}{Autoregressive Moving Average}
  \acro{ARQ}{\LU{A}{a}utomatic \LU{R}{r}epeat \LU{R}{r}equest}
  \acro{ATES}{Adaptive Throughput-based Efficiency-Satisfaction Trade-Off}
  \acro{AWGN}{additive white Gaussian noise}
  \acro{BAA}{\LU{B}{b}roadband \LU{A}{a}nalog \LU{A}{a}ggregation}
  \acro{BB}{Branch and Bound}
  \acro{BCD}{block coordinate descent}
  \acro{BD}{Block Diagonalization}
  \acro{BER}{Bit Error Rate}
  \acro{BF}{Best Fit}
  \acro{BFD}{bidirectional full duplex}
  \acro{BLER}{BLock Error Rate}
  \acro{BPC}{Binary Power Control}
  \acro{BPSK}{Binary Phase-Shift Keying}
  \acro{BRA}{Balanced Random Allocation}
  \acro{BS}{base station}
  \acro{BSUM}{block successive upper-bound minimization}
  \acro{CAP}{Combinatorial Allocation Problem}
  \acro{CAPEX}{Capital Expenditure}
  \acro{CBF}{Coordinated Beamforming}
  \acro{CBR}{Constant Bit Rate}
  \acro{CBS}{Class Based Scheduling}
  \acro{CC}{Congestion Control}
  \acro{CDF}{Cumulative Distribution Function}
  \acro{CDMA}{Code-Division Multiple Access}
  \acro{CE}{\LU{C}{c}hannel \LU{E}{e}stimation}
  \acro{CFO}{carrier frequency offset}
  \acro{CL}{Closed Loop}
  \acro{CLPC}{Closed Loop Power Control}
  \acro{CML}{centralized machine learning}
  \acro{CNR}{Channel-to-Noise Ratio}
  \acro{CNN}{\LU{C}{c}onvolutional \LU{N}{n}eural \LU{N}{n}etwork}
  \acro{CP}{computation point}
  \acro{CPA}{Cellular Protection Algorithm}
  \acro{CPICH}{Common Pilot Channel}
  \acro{CoCoA}{\LU{C}{c}ommunication efficient distributed dual \LU{C}{c}oordinate \LU{A}{a}scent}
  \acro{CoMAC}{\LU{C}{c}omputation over \LU{M}{m}ultiple-\LU{A}{a}ccess \LU{C}{c}hannels}
  \acro{CoMP}{Coordinated Multi-Point}
  \acro{CQI}{Channel Quality Indicator}
  \acro{CRM}{Constrained Rate Maximization}
	\acro{CRN}{Cognitive Radio Network}
  \acro{CS}{Coordinated Scheduling}
  \acro{CSI}{\LU{C}{c}hannel \LU{S}{s}tate \LU{I}{i}nformation}
  \acro{CSMA}{\LU{C}{c}arrier \LU{S}{s}ense \LU{M}{m}ultiple \LU{A}{a}ccess}
  \acro{CUE}{Cellular User Equipment}
  \acro{D2D}{device-to-device}
  \acro{DAC}{digital-to-analog converter}
  \acro{DC}{direct current}
  \acro{DCA}{Dynamic Channel Allocation}
  \acro{DE}{Differential Evolution}
  \acro{DFT}{Discrete Fourier Transform}
  \acro{DIST}{Distance}
  \acro{DL}{downlink}
  \acro{DMA}{Double Moving Average}
  \acro{DML}{Distributed Machine Learning}
  \acro{DMRS}{demodulation reference signal}
  \acro{D2DM}{D2D Mode}
  \acro{DMS}{D2D Mode Selection}
  \acro{DPC}{Dirty Paper Coding}
  \acro{DRA}{Dynamic Resource Assignment}
  \acro{DSA}{Dynamic Spectrum Access}
  \acro{DSGD}{\LU{D}{d}istributed \LU{S}{s}tochastic \LU{G}{g}radient \LU{D}{d}escent}
  \acro{DSM}{Delay-based Satisfaction Maximization}
  \acro{ECC}{Electronic Communications Committee}
  \acro{EFLC}{Error Feedback Based Load Control}
  \acro{EI}{Efficiency Indicator}
  \acro{eNB}{Evolved Node B}
  \acro{EPA}{Equal Power Allocation}
  \acro{EPC}{Evolved Packet Core}
  \acro{EPS}{Evolved Packet System}
  \acro{E-UTRAN}{Evolved Universal Terrestrial Radio Access Network}
  \acro{ES}{Exhaustive Search}
  \acro{FC}{\LU{F}{f}usion \LU{C}{c}enter}
  \acro{FD}{\LU{F}{f}ederated \LU{D}{d}istillation}
  \acro{FDD}{frequency divisionov duplex}
  \acro{FDM}{Frequency Division Multiplexing}
  \acro{FDMA}{\LU{F}{f}requency \LU{D}{d}ivision \LU{M}{m}ultiple \LU{A}{a}ccess}
  \acro{FedAvg}{\LU{F}{f}ederated \LU{A}{a}veraging}
  \acro{FER}{Frame Erasure Rate}
  \acro{FF}{Fast Fading}
  \acro{FL}{federated learning}
  \acro{FOMs}{first-order methods}
  \acro{FPP-SCA}{Feasible Point Pursuit-Successive Convex Approximation}
  \acro{FSB}{Fixed Switched Beamforming}
  \acro{FST}{Fixed SNR Target}
  \acro{FTP}{File Transfer Protocol}
  \acro{GA}{Genetic Algorithm}
  \acro{GBR}{Guaranteed Bit Rate}
  \acro{GLR}{Gain to Leakage Ratio}
  \acro{GOS}{Generated Orthogonal Sequence}
  \acro{GPL}{GNU General Public License}
  \acro{GRP}{Grouping}
  \acro{HARQ}{Hybrid Automatic Repeat Request}
  \acro{HD}{half-duplex}
  \acro{HMS}{Harmonic Mode Selection}
  \acro{HOL}{Head Of Line}
  \acro{HSDPA}{High-Speed Downlink Packet Access}
  \acro{HSPA}{High Speed Packet Access}
  \acro{HTTP}{HyperText Transfer Protocol}
  \acro{ICMP}{Internet Control Message Protocol}
  \acro{ICI}{Intercell Interference}
  \acro{ID}{Identification}
  \acro{IETF}{Internet Engineering Task Force}
  \acro{ILP}{Integer Linear Program}
  \acro{JRAPAP}{Joint RB Assignment and Power Allocation Problem}
  \acro{UID}{Unique Identification}
  \acro{IID}{\LU{I}{i}ndependent and \LU{I}{i}dentically \LU{D}{d}istributed}
  \acro{IIR}{Infinite Impulse Response}
  \acro{ILP}{Integer Linear Problem}
  \acro{IMT}{International Mobile Telecommunications}
  \acro{INV}{Inverted Norm-based Grouping}
  \acro{IoT}{Internet of Things}
  \acro{IP}{Integer Programming}
  \acro{IPMs}{interior-point methods}
  \acro{IPv6}{Internet Protocol Version 6}
  \acro{ISD}{Inter-Site Distance}
  \acro{ISI}{Inter Symbol Interference}
  \acro{ITU}{International Telecommunication Union}
  \acro{JAFM}{joint assignment and fairness maximization}
  \acro{JAFMA}{joint assignment and fairness maximization algorithm}
  \acro{JOAS}{Joint Opportunistic Assignment and Scheduling}
  \acro{JOS}{Joint Opportunistic Scheduling}
  \acro{JP}{Joint Processing}
	\acro{JS}{Jump-Stay}
  \acro{KKT}{Karush-Kuhn-Tucker}
  \acro{L3}{Layer-3}
  \acro{LAC}{Link Admission Control}
  \acro{LA}{Link Adaptation}
  \acro{LC}{Load Control}
  \acro{LDC}{\LU{L}{l}earning-\LU{D}{d}riven \LU{C}{c}ommunication}
  \acro{LOS}{line of sight}
  \acro{LP}{Linear Programming}
  \acro{LTE}{Long Term Evolution}
	\acro{LTE-A}{\ac{LTE}-Advanced}
  \acro{LTE-Advanced}{Long Term Evolution Advanced}
  \acro{LRA}{Low Rank Approximation}
  \acro{M2M}{Machine-to-Machine}
  \acro{MAC}{multiple access channel}
  \acro{MAP}{maximum a posterior}
  \acro{MANET}{Mobile Ad hoc Network}
  \acro{MC}{Modular Clock}
  \acro{MCS}{Modulation and Coding Scheme}
  \acro{MDB}{Measured Delay Based}
  \acro{MDI}{Minimum D2D Interference}
  \acro{MF}{Matched Filter}
  \acro{MG}{Maximum Gain}
  \acro{MH}{Multi-Hop}
  \acro{MIMO}{\LU{M}{m}ultiple \LU{I}{i}nput \LU{M}{m}ultiple \LU{O}{o}utput}
  \acro{MINLP}{mixed integer nonlinear programming}
  \acro{MIP}{mixed integer programming}
  \acro{MISO}{multiple input single output}
  \acro{ML}{machine learning}
  \acro{MLE}{maximum likelihood estimator}
  \acro{MLWDF}{Modified Largest Weighted Delay First}
  \acro{MME}{Mobility Management Entity}
  \acro{MMSE}{minimum mean squared error}
  \acro{MOS}{Mean Opinion Score}
  \acro{MPF}{Multicarrier Proportional Fair}
  \acro{MRA}{Maximum Rate Allocation}
  \acro{MR}{Maximum Rate}
  \acro{MRC}{Maximum Ratio Combining}
  \acro{MRT}{Maximum Ratio Transmission}
  \acro{MRUS}{Maximum Rate with User Satisfaction}
  \acro{MS}{Mode Selection}
  \acro{MSE}{\LU{M}{m}ean \LU{S}{s}quared \LU{E}{e}rror}
  \acro{MSI}{Multi-Stream Interference}
  \acro{MTC}{Machine-Type Communication}
  \acro{MTSI}{Multimedia Telephony Services over IMS}
  \acro{MTSM}{Modified Throughput-based Satisfaction Maximization}
  \acro{MU-MIMO}{Multi-User Multiple Input Multiple Output}
  \acro{MU}{Multi-User}
  \acro{NAS}{Non-Access Stratum}
  \acro{NB}{Node B}
	\acro{NCL}{Neighbor Cell List}
  \acro{NLP}{Nonlinear Programming}
  \acro{NLOS}{non-line of sight}
  \acro{NMSE}{normalized mean square error}
  \acro{NOMA}{\LU{N}{n}on-\LU{O}{o}rthogonal \LU{M}{m}ultiple \LU{A}{a}ccess}
  \acro{NORM}{Normalized Projection-based Grouping}
  \acro{NP}{non-polynomial time}
  \acro{NRT}{Non-Real Time}
  \acro{NSPS}{National Security and Public Safety Services}
  \acro{O2I}{Outdoor to Indoor}
  \acro{OFDMA}{\LU{O}{o}rthogonal \LU{F}{f}requency \LU{D}{d}ivision \LU{M}{m}ultiple \LU{A}{a}ccess}
  \acro{OFDM}{Orthogonal Frequency Division Multiplexing}
  \acro{OFPC}{Open Loop with Fractional Path Loss Compensation}
	\acro{O2I}{Outdoor-to-Indoor}
  \acro{OL}{Open Loop}
  \acro{OLPC}{Open-Loop Power Control}
  \acro{OL-PC}{Open-Loop Power Control}
  \acro{OPEX}{Operational Expenditure}
  \acro{ORB}{Orthogonal Random Beamforming}
  \acro{JO-PF}{Joint Opportunistic Proportional Fair}
  \acro{OSI}{Open Systems Interconnection}
  \acro{PAIR}{D2D Pair Gain-based Grouping}
  \acro{PAPR}{Peak-to-Average Power Ratio}
  \acro{P2P}{Peer-to-Peer}
  \acro{PC}{Power Control}
  \acro{PCI}{Physical Cell ID}
  \acro{PDCCH}{physical downlink control channel}
  \acro{PDD}{penalty dual decomposition}
  \acro{PDF}{Probability Density Function}
  \acro{PER}{Packet Error Rate}
  \acro{PF}{Proportional Fair}
  \acro{P-GW}{Packet Data Network Gateway}
  \acro{PL}{Pathloss}
  \acro{RLT}{reformulation linearization technique}
  \acro{PRB}{Physical Resource Block}
  \acro{PROJ}{Projection-based Grouping}
  \acro{ProSe}{Proximity Services}
  \acro{PS}{\LU{P}{p}arameter \LU{S}{s}erver}
  \acro{PSO}{Particle Swarm Optimization}
  \acro{PUCCH}{physical uplink control channel}
  \acro{PZF}{Projected Zero-Forcing}
  \acro{QAM}{quadrature amplitude modulation}
  \acro{QoS}{quality of service}
  \acro{QPSK}{quadrature phase shift keying}
  \acro{QCQP}{quadratically constrained quadratic programming}
  \acro{RAISES}{Reallocation-based Assignment for Improved Spectral Efficiency and Satisfaction}
  \acro{RAN}{Radio Access Network}
  \acro{RA}{Resource Allocation}
  \acro{RAT}{Radio Access Technology}
  \acro{RATE}{Rate-based}
  \acro{RB}{resource block}
  \acro{RBG}{Resource Block Group}
  \acro{REF}{Reference Grouping}
  \acro{ReLU}{rectified linear unit}
  \acro{ReMAC}{repetition for multiple access computing}
  \acro{RF}{radio frequency}
  \acro{RLC}{Radio Link Control}
  \acro{RM}{Rate Maximization}
  \acro{RNC}{Radio Network Controller}
  \acro{RND}{Random Grouping}
  \acro{RRA}{Radio Resource Allocation}
  \acro{RRM}{\LU{R}{r}adio \LU{R}{r}esource \LU{M}{m}anagement}
  \acro{RSCP}{Received Signal Code Power}
  \acro{RSRP}{reference signal receive power}
  \acro{RSRQ}{Reference Signal Receive Quality}
  \acro{RR}{Round Robin}
  \acro{RRC}{Radio Resource Control}
  \acro{RSSI}{received signal strength indicator}
  \acro{RT}{Real Time}
  \acro{RU}{Resource Unit}
  \acro{RUNE}{RUdimentary Network Emulator}
  \acro{RV}{Random Variable}
  \acro{SAA}{Small Argument Approximation}
  \acro{SAC}{Session Admission Control}
  \acro{SCA}{successive convex approximation}
  \acro{SCM}{Spatial Channel Model}
  \acro{SC-FDMA}{Single Carrier - Frequency Division Multiple Access}
  \acro{SD}{subgradient descent}
  \acro{S-D}{Source-Destination}
  \acro{SDPC}{Soft Dropping Power Control}
  \acro{SDMA}{Space-Division Multiple Access}
  \acro{SDR}{semidefinite relaxation}
  \acro{SDP}{semidefinite programming}
  \acro{SeMAC}{sequential modulation for AirComp}
  \acro{SeMAC-PA}{SeMAC with power adaptation}
  \acro{SER}{Symbol Error Rate}
  \acro{SES}{Simple Exponential Smoothing}
  \acro{S-GW}{Serving Gateway}
  \acro{SGD}{\LU{S}{s}tochastic \LU{G}{g}radient \LU{D}{d}escent}  
  \acro{SINR}{signal-to-interference-plus-noise ratio}
  \acro{SI}{self-interference}
  \acro{SIP}{Session Initiation Protocol}
  \acro{SISO}{\LU{S}{s}ingle \LU{I}{i}nput \LU{S}{s}ingle \LU{O}{o}utput}
  \acro{SIMO}{Single Input Multiple Output}
  \acro{SIR}{Signal to Interference Ratio}
  \acro{SLNR}{Signal-to-Leakage-plus-Noise Ratio}
  \acro{SMA}{Simple Moving Average}
  \acro{SNR}{\LU{S}{s}ignal-to-\LU{N}{n}oise \LU{R}{r}atio}
  \acro{SORA}{Satisfaction Oriented Resource Allocation}
  \acro{SORA-NRT}{Satisfaction-Oriented Resource Allocation for Non-Real Time Services}
  \acro{SORA-RT}{Satisfaction-Oriented Resource Allocation for Real Time Services}
  \acro{SPF}{Single-Carrier Proportional Fair}
  \acro{SRA}{Sequential Removal Algorithm}
  \acro{SRS}{sounding reference signal}
  \acro{SSD}{stochastic subgradient descent}
  \acro{SU-MIMO}{Single-User Multiple Input Multiple Output}
  \acro{SU}{Single-User}
  \acro{SVD}{singular value decomposition}
  \acro{SVM}{\LU{S}{s}upport \LU{V}{v}ector \LU{M}{m}achine}
  \acro{TCP}{Transmission Control Protocol}
  \acro{TDD}{time division duplex}
  \acro{TDMA}{\LU{T}{t}ime \LU{D}{d}ivision \LU{M}{m}ultiple \LU{A}{a}ccess}
  \acro{TNFD}{three node full duplex}
  \acro{TETRA}{Terrestrial Trunked Radio}
  \acro{TP}{Transmit Power}
  \acro{TPC}{Transmit Power Control}
  \acro{TTI}{transmission time interval}
  \acro{TTR}{Time-To-Rendezvous}
  \acro{TSM}{Throughput-based Satisfaction Maximization}
  \acro{TU}{Typical Urban}
  \acro{UE}{\LU{U}{u}ser \LU{E}{e}quipment}
  \acro{UEPS}{Urgency and Efficiency-based Packet Scheduling}
  \acro{UL}{uplink}
  \acro{UMTS}{Universal Mobile Telecommunications System}
  \acro{URI}{Uniform Resource Identifier}
  \acro{URM}{Unconstrained Rate Maximization}
  \acro{VR}{Virtual Resource}
  \acro{VoIP}{Voice over IP}
  \acro{WAN}{Wireless Access Network}
  \acro{WCDMA}{Wideband Code Division Multiple Access}
  \acro{WF}{Water-filling}
  \acro{WiMAX}{Worldwide Interoperability for Microwave Access}
  \acro{WINNER}{Wireless World Initiative New Radio}
  \acro{WLAN}{Wireless Local Area Network}
  \acro{WMMSE}{weighted minimum mean square error}
  \acro{WMPF}{Weighted Multicarrier Proportional Fair}
  \acro{WPF}{Weighted Proportional Fair}
  \acro{WSN}{Wireless Sensor Network}
  \acro{WWW}{World Wide Web}
  \acro{XIXO}{(Single or Multiple) Input (Single or Multiple) Output}
  \acro{ZF}{Zero-Forcing}
  \acro{ZMCSCG}{Zero Mean Circularly Symmetric Complex Gaussian}
\end{acronym}

\title{Multi-Symbol Digital AirComp via \\ Modulation Design and Power Adaptation
}

\author{Xiaojing Yan,~\IEEEmembership{Member,~IEEE}, Saeed Razavikia,~\IEEEmembership{Member,~IEEE}, Carlo Fischione,~\IEEEmembership{Fellow,~IEEE}\\
 	\thanks{All the authors are with the School of Electrical Engineering and Computer Science KTH Royal Institute of Technology, Stockholm, Sweden (e-mail: \{xiay,sraz,carlofi\}@kth.se). C. Fischione is also with Digital Futures of KTH. }
   
}

\newtheorem{theorem}{Theorem}
\newtheorem{prop}{Proposition}
\newtheorem{lem}{Lemma}
\newtheorem{rem}{Remark}

\maketitle
\thispagestyle{empty}
\pagestyle{empty}

\begin{abstract}

Recently, \ac{AirComp} leverages the superposition property of wireless channels to enable efficient function computation over a \ac{MAC}. {\color{black}However, existing digital AirComp methods either rely on single-symbol modulation, which limits flexibility and robustness, or on multi-symbol extensions that suffer from high complexity or approximation errors. To overcome these limitations, we propose a new multi-symbol modulation framework, termed \ac{SeMAC}, which encodes each input into a sequence of symbols with distinct constellation diagrams across multiple time slots. This approach increases design flexibility and robustness against channel noise. Specifically, the modulation design is formulated as a non-convex optimization problem and efficiently solved through a \ac{SCA} combined with \ac{SSD}. For fixed modulation formats, we further develop \ac{SeMAC-PA} to adjusts transmit power and phase while preserving the modulation structure. Notably, numerical results show that \ac{SeMAC} improves computation accuracy by up to~14 dB compared to the existing methods for computing nonlinear functions such as the product function.}

\end{abstract}


\section{Introduction}

\acresetall

With the widespread deployment of \ac{IoT}, cellular networks are shifting from “connected everything” in 5G to “connected intelligence” in 6G systems~\cite{letaief2019roadmap}. This shift is driven by the growing need to process distributed data at the network edge devices~\cite{wang2023road}. In this context, \ac{AirComp} has emerged as a promising technique that leverages the signal superposition property to compute functions directly during transmission~\cite{csahin2023survey}. 
{\color{black}In particular, \ac{AirComp} supports 6G intelligence applications such as collaborative sensing and \ac{AirFL}, where local model updates are efficiently aggregated across devices to enable global model training~\cite{zhu2018mimo}}.

Analog \ac{AirComp} relies on uncoded transmission, making it vulnerable to noise and fading, and incompatible with modern digital infrastructures. To address these limitations, several digital \ac{AirComp} methods have been developed,
such as one-bit broadband digital aggregation (OBDA), majority-vote frequency-shift keying (FSK), and their asynchronous and non-coherent extensions. However, these methods are often limited to fewer number of transmitters or specific functions~\cite{csahin2023survey}. To provide a more general solution, ChannelComp~\cite{saeed2023ChannelComp} was introduced to enable arbitrary function computation over a \ac{MAC} by designing digital modulation schemes via optimization, and it adopts a single-symbol modulation design where each input is mapped to a single modulated symbol.

{\color{black}While ChannelComp is compatible with
existing digital communication systems, its single-symbol design limits computation accuracy under noisy conditions and its \ac{SDR} solution involves high optimization complexity. As a result, recent works have focused on multi-symbol modulation designs, which provide additional redundancy and improve robustness~\cite{yan2025remac,yan2024novel,liu2025digital}. For instance,
\ac{ReMAC}~\cite{yan2025remac} enhances computation robustness by selectively repeating modulated symbols across multiple time slots, but this repetition coding design further increases computational complexity. Bit-Slicing~\cite{liu2025digital} offers a lower computational complexity design by partitioning bit sequences into slices and mapping each sliced integer to square \ac{QAM} symbols for transmission, yet it is mainly effective for the sum function and suffers from approximation errors when extended to nonlinear computations. }

{\color{black}Motivated by these limitations, we propose \ac{SeMAC}, a new multi-symbol modulation framework that encodes each input into a sequence of symbols using distinct constellation diagrams.
As opposed to \ac{ReMAC} which uses an identical constellation pattern across all time slots, \ac{SeMAC} leverages constellation diversity to achieve higher computation accuracy. The resulting modulation design problem is formulated as a non-convex \ac{QCQP}, which is NP-hard and computationally demanding~\cite{luo2010semidefinite}. To address this complexity challenge, we develop an iterative algorithm based on \ac{SCA} combined with \ac{SSD}, which achieves lower complexity than the standard \ac{SDR} method. 
In addition, for scenarios where the modulation format is fixed, we propose \ac{SeMAC-PA}, which adapts the transmit amplitude and phase of the existing modulation patterns, thereby extending the applicability of our approach to hardware-constrained systems.
Numerical results show that \ac{SeMAC} enables robust computation over a noisy \ac{MAC} and reduces the computation error by up to 14~dB compared to ReMAC, particularly for the product function.}

\input{Figures/Fig-system}

\section{Preliminaries} \label{sec:Preliminaries}


\subsection{Signal Model}  \label{subsec:signal_model}

Consider a network with $K$ single-antenna nodes transmitting data to a \ac{CP} over a shared channel. Each node $k$ sends a data value $x_k \in \mathbb{F}_Q $, where $\mathbb{F}_Q$ denotes a finite field with $Q=2^b$ elements, where $x_k$ is represented using $b$ bits. The goal of the \ac{CP} is to compute any finite-valued function $f(x_1,\ldots,x_K)$ over the \ac{MAC}s. Each node encodes its data value $x_k$ into a modulated signal $\vec{x}_k$ through an encoder $\mathcal{E}_k$, such that $\vec{x}_k = \mathcal{E}_k(x_k)\in \mathbb{C}$. We assume perfect synchronization of carrier frequency and symbol timing among all the nodes and the \ac{CP}\footnote{{\color{black} In practical systems, techniques such as frame timing and carrier frequency
offset estimation can be applied for achieving synchronization~\cite{guo2021over}.}}, and then the \ac{CP} receives the superimposed signal as
\begin{align}
\label{eq:receivedsignal}
\vec{y} = \sum\nolimits_{k=1}^{K} h_k p_k \vec{x}_k + \vec{z},
\end{align}  
where \( h_k \) and \( p_k \) denote the channel coefficient and transmit power for node \( k \), and \( \vec{z} \sim \mathcal{N}(0, \sigma_z^2) \) represents the additive white Gaussian noise. 

Assuming the availability of perfect \ac{CSI}, each node applies optimal power control by inverting the channel~\cite{cao2019optimal}, given by $p_k = h_k^*/|h_k|^2$. In this regard, we can compensate for channel distortion and ensure coherent signal alignment at the \ac{CP}.
With channel inversion applied, and ignoring the effect of the noise, the received signal $\vec{r}:=\sum\nolimits_{k=1}^K \vec{x}_k$ yields a finite constellation diagram. Then, the \ac{CP} applies a tabular function $\mathcal{T}(\cdot)$ to map each constellation point to its corresponding function output.

\subsection{Overlap Avoidance} \label{subsec:overlap avoidance}

Consider a noiseless \ac{MAC}, and let $f^{(i)}$ and $f^{(j)}$ be two distinct function values corresponding to input sets $x_1^{(i)}, \dots, x_K^{(i)}$ and $x_1^{(j)}, \dots, x_K^{(j)}$, respectively. The aggregated constellation points are given by $\vec{v}^{(i)} := \sum\nolimits_{k=1}^{K} \vec{x}_{k}^{(i)}$ and $
\vec{v}^{(j)} := \sum\nolimits_{k=1}^{K} \vec{x}_{k}^{(j)}$. 
To ensure valid computation, destructive overlap between $\vec{v}^{(i)}$ and $\vec{v}^{(j)}$ must be avoided, so that the tabular function $\mathcal{T}(\cdot)$ can uniquely map the constellation point $\vec{v}^{(i)}$ to its corresponding function value $f^{(i)}$. This leads to the following overlap avoidance constraint by a smooth condition~\cite{saeed2023ChannelComp}: 
\begin{align}
\label{eq:CompCond_smooth}
|\vec{v}^{(i)} - \vec{v}^{(j)}| \geq \epsilon |f^{(i)} - f^{(j)}|^2, \quad \forall (i,j) \in \mathcal{M},
\end{align}  
{\color{black}where $\mathcal{M}:=\{(i,j)\in [M]^2|i<j\}$ and \( \epsilon>0\) is a positive constant. Specifically, $M$ is the cardinality of the function range, and $\tilde{M}$ is the number of index pair in set $\mathcal{M}$}.

\section{System Model}\label{sec:Systemmodel}



\subsection{Multi-Symbol Transmission} \label{subsec:transmitter}

Extending the signal model in~\ref{subsec:signal_model}, \ac{SeMAC} enables multi-symbol transmission by encoding each input value $x_k$ at node $k$ into a sequence of $L$ modulated symbols using a modulation encoder, expressed as $\mathcal{E}_k(x_k) = [\vec{x}_{k,1}, \dots, \vec{x}_{k,L}] \in \mathbb{C}^{1 \times L}$. These symbols are then transmitted sequentially over $L$ consecutive time slots, with $\vec{x}_{k,\ell}$ corresponding to time slot $\ell$.
Consequently, the superimposed signal received at the \ac{CP} in each time slot is given by:
\begin{equation}
\label{eq:fading_multislot}
\vec{y}_{\ell} = \sum\nolimits_{k=1}^{K} h_{k,\ell} p_{k,\ell} \vec{x}_{k,\ell} + \vec{\tilde{z}}_{\ell}, \quad \forall~\ell \in [L],
\end{equation}
where \( h_{k,\ell} \) and \( p_{k,\ell} \) denote the channel coefficient and transmit power of node \( k \) at time slot \( \ell \), and \( \vec{\tilde{z}}_{\ell} \sim \mathcal{N}(0, \sigma_{z}^2) \) is the additive white Gaussian noise at time slot $\ell$. Similarly, following the optimal power control policy~\cite{cao2019optimal}, the received signal can be written as:
\begin{equation}
\label{eq:nofading_multislot}
\vec{y}_{\ell} = \sum\nolimits_{k=1}^{K}\vec{x}_{k,\ell} + \vec{\tilde{z}}_{\ell}, \quad \forall~\ell \in [L].
\end{equation}

For each node $k$, we define the modulation matrix as $\bm{X}_k \in \mathbb{C}^{Q \times L}$, where $[\bm{X}_k]_{(q,\ell)} := \vec{x}_{k,\ell}^{(q)}$ denotes the signal transmitted at time slot $\ell$ when the input is $x_k^{(q)} \in \mathbb{F}_Q$. In this matrix, each row corresponds to an input value and contains its encoded modulation sequence $\bm{x}_k^{(q)} := \mathcal{E}_k(x_k^{(q)}) = [\vec{x}_{k,1}^{(q)}, \ldots, \vec{x}_{k,L}^{(q)}]$. Meanwhile, each column $\bm{x}_{k,\ell} := [\vec{x}_{k,\ell}^{(1)}, \ldots, \vec{x}_{k,\ell}^{(Q)}]^{\mathsf{T}} \in \mathbb{C}^{Q \times 1}$ defines the modulation pattern used at time slot $\ell$, containing the $Q$ complex symbols assigned to all possible input values. Without loss of generality,  each modulation vector follows a unit norm, i.e., $\|\bm{x}_{k,\ell}\|_2^2 \leq 1$, $\forall k \in [K]$ and $\forall \ell \in [L]$.

Furthermore, we construct the modulation matrix $\bm{X} := [\bm{X}_1^{\mathsf{T}}, \dots, \bm{X}_K^{\mathsf{T}}]^{\mathsf{T}} \in \mathbb{C}^{N \times L}$, where $N := QK$, by vertically stacking the modulation matrices of all $K$ nodes. Give the input combination $x_1^{(i)}, \dots, x_K^{(i)}$, we define a binary vector $\bm{a}_i \in \{0,1\}^N$ which consists of $K$ one-hot blocks of length $Q$, each selecting the encoded modulation symbol at one node. Therefore, the resultant modulation sequence associated with the function output $f^{(i)}$ is given by $\bm{v}^{(i)} = \bm{a}_i \bm{X} \in \mathbb{C}^{L}$.

\subsection{Multi-Symbol Estimation and Decoding}

After receiving the aggregated signal over the \ac{MAC}, the \ac{CP} needs to estimate the transmitted constellation sequence, and then maps the estimated sequence to the output of
function $f$.
More precisely, in each time slot $\ell$, a maximum likelihood estimator partitions the constellation diagram, which consists of all possible constellation points \( \{\vec{v}_\ell^{(1)}, \ldots, \vec{v}_\ell^{(M)}\} \) along with their corresponding Voronoi cells \( \{\mathcal{V}_{1,\ell}, \ldots, \mathcal{V}_{M,\ell}\} \), and maps the received symbol $\vec{y}_{\ell}$ to the closest constellation point as $\vec{v}_{\ell} = \arg \min_{i} \| \vec{y}_{\ell} - \vec{v}^{(i)}_{\ell} \|_2^2$.
Then, the recovered function output $\hat{f}$ is computed via the tabular function $\hat{f} = \sum\nolimits_{j=1}^{M} \mathcal{T}_j(\bm{v})$, where  $\mathcal{T}_j(\cdot)$ is an indicator function:
\begin{align}
\mathcal{T}_{j}(\bm{v}) := \begin{cases}
    \hat{f}^{(j)}, & \quad \text{if } \vec{v}_{\ell} \in \mathcal{V}_{j,\ell},\text{ } \forall \ell \in [L], \\
0, & \quad \textrm{otherwise}. 
\end{cases}    
\end{align}

To ensure error-free computation, the tabular function must uniquely associate each received constellation sequence with its corresponding function value. In contrast to ChannelComp, where decoding relies on individual symbols, correct recovery in \ac{SeMAC} depends on the received modulation sequence. Thus, any two sequences $\bm{v}_i$ and $\bm{v}_j$ must be separated in Euclidean space when their function outputs $f^{(i)}$ and $f^{(j)}$ are different. This requirement leads to a generalized overlap avoidance constraint for designing multi-symbol modulation:
\begin{align}
\label{eq:CompCond_multislot}
\| (\bm{a}_{i} - \bm{a}_{j})^{\mathsf{T}}\bm{X}\|_2^2 \geq \epsilon |f^{(i)} - f^{(j)}|, \quad \forall (i,j) \in \mathcal{M},
\end{align}
where \( \|\cdot\|_2 \) denotes the \( \ell_2 \)-norm. Note that if the condition in~\eqref{eq:CompCond_multislot} is not satisfied, the corresponding Voronio regions are merged, and we consider the average of the associated outputs as the final function output~\cite{saeed2023ChannelComp}.
The overall \ac{SeMAC} process is shown in Figure~\ref{fig:digital_aircomp}.

\section{Modulation Design and Power Adaptation}

In this section, we consider a scenario where the network allocates up to $L$ time slots, with $Q>L$. When modulation patterns are configurable, we employ \ac{SeMAC} to design the modulation. In contrast, when modulation is fixed, we adopt \ac{SeMAC-PA}, which adapts the transmit power and phase to reshape the received sequence of the constellation.

\subsection{Modulation Design With Channel Inversion}

In this subsection, the modulation scheme is designed under the assumption of channel inversion power control. 
To avoid constellation overlaps as in~\eqref{eq:CompCond_multislot}, we pose the following feasibility problem to enable robust computation, subject to the norm power constraint:
\begin{subequations}
\label{eq:smooth original problem}
\begin{align}
\nonumber
 \mathcal{P}_{0}:\quad& {\rm find} \quad \bm{X} \\ 
 \label{eq:original separation constraint}
 {\rm s.t.} \quad & {\rm Tr}(\bm{X}^{\mathsf{H}} \bm{D}_{ij} \bm{X}) \geq  \Delta f_{ij}, \quad \forall (i,j) \in \mathcal{M},  \\ \label{eq:original rank constraint}
\quad & \|\bm{x}_{k,\ell}\|_2^2 \leq 1,   \quad \forall k \in [K], \quad \forall \ell \in [L], 
\end{align}
\end{subequations}
where $\Delta f_{ij} := \epsilon |f^{(i)} - f^{(j)}|$ and $\bm{D}_{ij} := (\bm{a}_i - \bm{a}_j)(\bm{a}_i - \bm{a}_j)^\mathsf{T}$.
However, \( \mathcal{P}_0 \) is a non-convex \ac{QCQP}, and solving it is NP-hard~\cite{saeed2023ChannelComp}. {\color{black}A common approach to tackle such problems is the \ac{SDR}~\cite{yan2025remac}, which uses matrix lifting coupled with rank relaxation to relax the problem into a convex \ac{SDP}. While effective for small instances, \ac{SDR} consumes potentially high complexity for large $K$ and $Q$. 
To overcome this, we consider a scalable alternative based on \ac{SCA} to approximate the original problem via a sequence of convex subproblems~\cite{beck2010sequential}. Specifically, we first formulate the feasibility problem into an unconstrained minimization problem by hinge-loss penalites:
\begin{equation}
\label{eq:SCA_reformulation_set}
\mathcal{P}_1 : \quad \min_{\bm{X} \in \mathcal{X}} \quad 
\frac{1}{\tilde{M}}\sum\nolimits_{(i,j) \in \mathcal{M}} g_{ij}(\bm{X}),
\end{equation}
where $g_{ij}(\bm{X}):=\left[ \Delta f_{ij} - \operatorname{Tr}(\bm{X}^{\mathsf{H}} \bm{D}_{ij} \bm{X}) \right]^+$ penalizes constraint violations and $\mathcal{X}:=\{\|\bm{x}_{k,\ell}\|_2^2\leq 1,\forall k\in [K], \forall \ell \in [L]\}$ is a convex set of power-constrained vectors. Since $g_{ij}(\bm{X})$ is non-convex, we then approximate it by a convex majorizer at a reference point $\tilde{\bm{X}}^{(s)}$, leading to the following function:}
{\color{black}
\begin{align}
\label{eq:convex expansion} \nonumber
h_{ij}^{(s)}(\bm{X}):=&\left[\Delta f_{ij}-2\Re\{\operatorname{Tr}[(\tilde{\bm{X}}^{(s)})^{\mathsf{H}}\bm{D}_{ij}\bm{X}]\} \right.\\ 
&\left. +{\rm Tr}[(\tilde{\bm{X}}^{(s)})^{\mathsf{H}}\bm{D}_{ij}\tilde{\bm{X}}^{(s)}]\right]^+.
\end{align}
Hence, at \ac{SCA} iteration $s$, we solve the following convex subproblem $\mathcal{P}_2$, whose solution $\bm{\tilde{X}}^{(s+1)}$ serves as the reference point for the next iteration:
\begin{equation}
\label{eq:SCA_relaxed_reformulation}
\mathcal{P}_2 : \bm{\tilde{X}}^{(s+1)} \in \arg\min_{\bm{X} \in \mathcal{X}} 
\frac{1}{\tilde{M}}\sum\nolimits_{(i,j) \in \mathcal{M}} h_{ij}^{(s)}(\bm{X}).
\end{equation}
\begin{algorithm}[!t] 
\caption{{\color{black}\ac{SCA}-\ac{SSD} for solving Problem $\mathcal{P}_0$}}
\label{Alg:HIT}
\begin{algorithmic}[1] 
\STATE {\color{black}\textbf{Init:} \ac{SCA} iterations $S$, subgradient steps $T$, step sizes $\{\alpha_t\}_{t=0}^{T-1}$. Randomly generate an initial point $\bm{\tilde{X}}^{(0)} \in \mathcal{X}$}.
{\color{black}\FOR{$s=0,1,\dots,S$}
    \FOR{$t=0,1,\dots,T-1$}
      \STATE Sample a single pair $(i_t,j_t)$ uniformly from $\mathcal M$
         \STATE $\bm{G}^{(t)} = \partial h^{(s)}_{i_t j_t}\!\left(\bm{X}^{(t)}\right)$
      \STATE 
      $\bm{Y}^{(t+1)} = \bm{X}^{(t)} - \alpha_t\, \bm{G}^{(t)}$
      \STATE 
      $\bm{X}^{(t+1)} = \Pi_{\mathcal X}\!\left(\bm{Y}^{(t+1)}\right)$
    \ENDFOR
    \STATE Set $\bm{\tilde{X}}^{(s+1)}=\bm{X}^{(T)}$
\ENDFOR}

\STATE {\color{black}\textbf{Output:} $\hat {\bm{X}} \gets \bm{\tilde{X}}^{(S+1)}$}
\end{algorithmic}
\end{algorithm}

Instead of solving $\mathcal{P}_2$ by interior-point methods, which consumes high computation cost, we can determine an approximate solution by using the \ac{SSD} approach.
Specifically, at each step $t$, we randomly draw the index pair $(i_t,j_t)$ from set $\mathcal{M}$ and apply the following update rule:
\begin{subequations}
\label{eq:stochastic update rule}
\begin{align}
\label{eq:stochastic update rule 1}
\bm{Y}^{(t+1)} &= \bm{X}^{(t)} - \alpha_t \cdot \partial h_{i_tj_t}^{(s)}(\bm{X}^{(t)}), \\
\bm{X}^{(t+1)} &= \Pi_{\mathcal{X}}(\bm{Y}^{(t+1)}).
\end{align}
\end{subequations}
where $\alpha_t$ is the step size and the subgradient $\partial h_{ij}^{(s)}(\bm{X}^{(t)})$ at the point $\bm{X}^{(t)}$ is computed as:
\begin{align}
\label{eq:subgradient rule} 
\partial h_{ij}^{(s)}(\bm{X}^{(t)}) = \begin{cases}
   -2 \bm{D}_{ij} \bm{\tilde{X}}^{(s)}, & \text{if } h_{ij}(\bm{X}^{(t)}) >0 \\
\bm0, &  \textrm{otherwise}. 
\end{cases}
\end{align}
Moreover, $\Pi_{\mathcal{X}}(\cdot)$ denotes the projection onto $\mathcal{X}$ given by:
\begin{equation}
\label{eq:projection rule}   
\bm{x}_{k,\ell} :=
\begin{cases}
\bm{y}_{k,\ell}, & \text{if } \|\bm{y}_{k,\ell}\|_2 \le 1, \\
\frac{\bm{y}_{k,\ell}}{\|\bm{y}_{k,\ell}\|_2}, & \text{otherwise}.
\end{cases}
\end{equation}

The complete process is summarized in Algorithm~\ref{Alg:HIT}, which integrates the \ac{SCA}  with \ac{SSD} updates. }

\begin{rem}
\label{rem:convergence and complexity analysis}

{\color{black}Note that solving the standard \ac{SDR} problems require $\mathcal{O}(\tilde M^4N^{0.5}L^{0.5})$ operations due to matrix lifting~\cite{yan2025remac}, which becomes computationally demanding for large $K$ or $Q$. In contrast, the proposed SCA–SSD algorithm samples only one constraint per update, leading to per-iteration complexity $\mathcal{O}(NL)$ and total cost $\mathcal{O}(STNL)$. This linear scaling avoids dependence on $\tilde M$ and thereby ensures scalability in larger networks. Furthermore, with a diminishing step-size schedule, the SSD updates follow the expected descent direction and the framework converges in expectation to a stationary point~\cite{konar2017first}.}

\end{rem}

\subsection{Power Adaptation of Existing Modulation}

Due to hardware constraints, modifying the existing modulation format at each device is often not feasible. Moreover, channel inversion is sensitive to deep fades and may lead to excessive power consumption under realistic fading conditions. To address this, we propose \ac{SeMAC-PA}, an extension of the E-ChannelComp framework~\cite{saeed2023ChannelComp}, to enable realible function computation with fixed modulation through power and phase adaptation across multiple time slots.

Given the signal model in~\eqref{eq:fading_multislot}, we define $\bm{p}_\ell := [p_{1,\ell}, \ldots, p_{K,\ell}]^{\mathsf{T}} \in \mathbb{C}^K$ and $\bm{h}_\ell := [h_{1,\ell}, \ldots, h_{K,\ell}]^{\mathsf{T}} \in \mathbb{C}^K$ as the transmit power vector and channel vector at time slot $\ell$, respectively. Let $\tilde{\bm{x}}_{\ell}$ denotes the modulation vector containing a fixed modulation pattern, i.e., \ac{QAM}. In the presence of fading, the received constellation point corresponding to the function output $f^{(i)}$ at time slot $\ell$ is then given by:
\begin{equation}
\label{eq:received_sequence_fading_power_control}
\tilde{\bm{v}}^{(i)}_{\ell} = \bm{a}_i^{\mathsf{T}} \left( \text{diag}(\bm{h}_\ell) \otimes \bm{I}_Q \right) \text{diag}(\tilde{\bm{x}}_{\ell}) \left( \bm{I}_K \otimes \mathds{1}_Q \right) \bm{p}_\ell,
\end{equation}
where \( \otimes \) denotes the Kronecker product. \( \bm{I}_Q \) and \( \bm{I}_K \) are identity matrices of size \( Q \times Q \) and \( K \times K \). \( \mathds{1}_Q \) is a \( Q \times 1 \) vector with all elements equal to one, and the operator \( \text{diag}(\cdot) \) constructs a diagonal matrix. For simplicity, we let $\bm{B}_{\ell}:=\left( \text{diag}(\bm{h}_{\ell}) \otimes \bm{I}_Q \right) \text{diag}(\tilde{\bm{x}}_{\ell})(\bm{I}_K\otimes \mathds{1}_Q)$. Accordingly, we formulate the following \ac{QCQP} to determine the minimum power vectors $\bm{p}_\ell$ that satisfiy the constraints in~\eqref{eq:CompCond_multislot}:
\begin{subequations}
\label{eq:power_minimization}
\begin{align}
\nonumber
\mathcal{P}_3 :\quad & \min_{\{\bm{p}_\ell\}\in \mathcal{P}} \quad \sum\nolimits_{\ell=1}^L \|\bm{p}_\ell\|_2^2 \\ 
{\rm s.t.} \quad & \sum\nolimits_{\ell=1}^L
\bm{p}_\ell^{\mathsf{H}} \bm{C}_{\ell}^{ij}\bm{p}_\ell
\geq \Delta f_{ij}, \quad \forall(i,j) \in \mathcal{M}, 
\end{align}
\end{subequations}
where \( \bm{C}_{\ell}^{ij} := \bm{B}_{\ell}^{\mathsf{H}} (\bm{a}_i - \bm{a}_j) (\bm{a}_i - \bm{a}_j)^{\mathsf{T}} \bm{B}_{\ell} \) and $\mathcal{P}:=\{\bm{p}_\ell \in \mathbb{C}^K, \ell \in [L]\}$. {\color{black}In a similar manner to $\mathcal{P}_0$, we translate the quadratic constraints in $\mathcal{P}_3$ into hinge-loss penalties:
\begin{equation}
\label{eq:hinge-loss penalty PA} \nonumber
u_{ij}(\{\bm{p}_\ell\})=\left[\Delta f_{ij} - \sum\nolimits_{\ell=1}^L \bm{p}_\ell^{\mathsf{H}}\bm{C}_\ell^{ij}\bm{p}_\ell \right]^+, \quad  \forall(i,j) \in \mathcal{M},
\end{equation}
which measure the constraint violations. 
At each \ac{SCA} iteration $s$, we linearize these non-convex terms at the reference points $\{\bm{\tilde{p}}_\ell^{(s)}\}$, leading to the following convex subproblem:
\begin{equation}
\label{eq:SCA_relaxed_reformulation_PA}
\mathcal{P}_4 : \min_{\{\bm{p}_\ell\}\in \mathcal{P}} 
\sum\nolimits_{\ell=1}^L \bm{p}_\ell^{\mathsf{H}}\bm{I}_K\bm{p}_\ell + \frac{1}{\tilde{M}}\sum_{(i,j) \in \mathcal{M}} w_{ij}^{(s)}(\{\bm{p}_\ell\}),
\end{equation}
where the convex majorizer is defined as:
\begin{align}
\label{eq:convex expansion_PA} \nonumber
w_{ij}^{(s)}(\{\bm{p}_\ell\}):=&\left[\Delta f_{ij}-\sum\nolimits_{\ell=1}^L 2\Re\{(\bm{\tilde{p}}_\ell^{(s)})^{\mathsf{H}}\bm{C}_{\ell}^{ij}\bm{p}_\ell\} \right.\\ 
&\left. + \sum\nolimits_{\ell=1}^L(\bm{\tilde{p}}_\ell^{(s)})^{\mathsf{H}}\bm{C}_{\ell}^{ij}\bm{\tilde{p}}_\ell^{(s)}\right]^+.
\end{align}

For the employment of \ac{SSD}, a subgradient of $w_{ij}^{(s)}(\{\bm{p}_\ell\})$ with respect to each variable $\bm{p}_\ell$ is given by
\begin{equation}
\label{eq:subgradient_w}
\partial_{\bm{p}_\ell} w_{ij}^{(s)}\left(\{\bm p_\ell\}\right)=
\begin{cases}
-2\bm C_{\ell}^{ij}\bm{\tilde p}_\ell^{(s)}, & \text{if } w_{ij}^{(s)}\left(\{\bm p_\ell\}\right) >0,\\
\bm 0, & \text{otherwise}.
\end{cases}
\end{equation}

Overall, the complete procedure for solving $\mathcal{P}_3$ is summarized in Algorithm~\ref{Alg:PA}.}

\begin{algorithm}[!t] 
\caption{\ac{SCA}-\ac{SSD} for solving Problem $\mathcal{P}_3$}
\label{Alg:PA}
\begin{algorithmic}[1] 
\STATE {\color{black}\textbf{Init:} \ac{SCA} iterations $S$, subgradient steps $T$, step sizes $\{\alpha_t\}_{t=0}^{T-1}$. Randomly generate initial points $\{\bm{\tilde{p}}_\ell^{(0)}\} \in \mathcal{P}$.}
{\color{black}\FOR{$s=0,1,\dots,S$}
    \FOR{$t=0,1,\dots,T-1$}
      \STATE Sample a single pair $(i_t,j_t)$ uniformly from $\mathcal M$
         \STATE $\bm{G}_\ell^{(t)} =  2\bm{p}_\ell^{(t)} + \partial_{\bm{p}_\ell} w_{ij}^{(s)}(\{\bm{p}_\ell^{(t)}\}), \forall \ell \in [L]$
      \STATE 
      $\bm{p}_\ell^{(t+1)} = \bm{p}_\ell^{(t)} - \alpha_t\, \bm{G}^{(t)}_\ell, \forall \ell \in [L]$
    \ENDFOR
    \STATE Set $\bm{\tilde{p}}_\ell^{(s+1)}=\bm{p}_\ell^{(T)}, \forall \ell \in [L]$
\ENDFOR}

\STATE {\color{black}\textbf{Output:} $\hat {\bm{p}_\ell} \gets \bm{\tilde{p}}_\ell^{(S+1)}, \forall \ell \in [L]$}
\end{algorithmic}
\end{algorithm}
\vspace{-1em}


\section{Numerical Experiments}\label{sec:Num}

In this section, we evaluate the performance of \ac{SeMAC} as well as its power adaptive scheme \ac{SeMAC-PA}, and make a comparison with ChannelComp, ReMAC and Bit-Slicing. Specifically, the performance is evaluated using the \ac{NMSE} metric, defined as: 
\begin{equation}
\text{\ac{NMSE}}:=\frac{\sum\nolimits_{j=1}^{N_s}|f^{(i)}-\hat{f}_j^{(i)}|^2}{N_s|f_{{\rm max}} - f_{{\rm min}}|^2},    
\end{equation}
where $N_s=100$ is the number of Monte Carlo trials, $f^{(i)}$
is the desired function value, $\hat{f}^{(i)}_j$ is the estimated value in the $j$-th Monte Carlo trial. $f_{{\rm max}}$ and 
$f_{{\rm min}}$ denote the maximum and minimum values of the function output, respectively. {\color{black}The fading coefficients \( h_{k,\ell} \) are modeled as independent complex Gaussian random variables, i.e., \( h_{k,\ell} \sim \mathcal{CN}(0, 1)\). For the \ac{SSD} updates, we adopt a diminishing step-size schedule $\alpha_t=\alpha_0 / \sqrt{t+1}$, where $\alpha_0=0.5$.}

\subsection{Performance of \ac{SeMAC}}

For the first experiment, we consider a network of $K=8$ nodes and evaluate the computation of the sum function $f=\sum_{k=1}^K x_k$ and the product function $f=\prod_{k=1}^K x_k$, where $x_k\in {1,2,3,4}$. The results in Figure~\ref{fig:comparison_K_and_L} show that the \ac{NMSE} consistently decreases as the \ac{SNR} increases. In addition, increasing the number of time slots from $L=1$ to $L=2$ improves the computation accuracy by providing greater flexibility in constellation design. 
{\color{black}Moreover, our proposed SCA–SSD method generates solutions that achieve lower \ac{NMSE} than the \ac{SDR}-based method, as it directly refines solutions via minimizing a hinge-loss penalization within the original problem space, thereby avoiding the relaxation errors introduced by lifting and randomization in \ac{SDR}.}


\begin{figure}[!t]
\centering
\begin{tikzpicture}
    \begin{axis}[
        xlabel = {SNR},
        ylabel = {NMSE},
        label style={font=\footnotesize},
        width=0.42\textwidth,
        height=5.6cm,
        xmin=10, xmax=35,
        ymin=0.000002, ymax=0.03,
        legend style={nodes={scale=0.65, transform shape}, at={(0.3,0.85)}},
        ticklabel style = {font=\footnotesize},
        legend pos=south west,
        ymajorgrids=true,
        xmajorgrids=true,
        grid style=dashed,
        grid=both,
        ymode = log,
        grid style={line width=.1pt, draw=gray!10},
        major grid style={line width=.2pt,draw=gray!30},
    ]
    \addplot[smooth,
             thin,
             dashed,
        color=chestnut,
        mark=*,
        mark options = {rotate = 180,solid},
        line width=0.75pt,
        mark size=1pt,
        ]
    table[x=SNR,y=sum]
    {Data/Sim_SCA.dat};
    \addplot[ smooth,
              thin,
              dashed,
            color=airforceblue,
            mark=+,
            mark options = {rotate = 45,solid},
            line width=0.75pt,
            mark size=2.5pt,
            ]
    table[x=SNR,y=sum2]
    {Data/Sim_SCA.dat};
    \addplot[ smooth,
             thin,
             dashed,
        color=cssgreen,
        mark=asterisk,
        mark options = {rotate = 180, solid},
        line width=0.75pt,
        mark size=2pt,
        ]
    table[x=SNR,y=sum2_SDR]
    {Data/Sim_SCA.dat};
    \addplot[ smooth,
             thin,
        color=chestnut,
        mark=square,
        line width=0.75pt,
        mark size=2pt,
        ]
    table[x=SNR,y=prod]
    {Data/Sim_SCA.dat};
    \addplot[ smooth,
             thin,
        color=airforceblue,
        mark=triangle,
        mark options = {rotate = 180, solid},
        line width=0.75pt,
        mark size=2.5pt,
        ]
    table[x=SNR,y=prod2]
    {Data/Sim_SCA.dat};
    \addplot[ smooth,
             thin,
        color=cssgreen,
        mark=triangle,
        line width=0.75pt,
        mark size=2.5pt,
        ]
    table[x=SNR,y=prod2_SDR]
    {Data/Sim_SCA.dat};
\legend{$\sum_k  L=1$ SCA-SSD,$\sum_k L=2$ SCA-SSD,$\sum_k L=2$ SDR,$\prod_k L=1$ SCA-SSD,$\prod_k L=2$ SCA-SSD,$\prod_k L=2$ SDR};
\end{axis}
\end{tikzpicture}
\vspace{-1em}
\caption{{\color{black}Performance of SeMAC in terms of NMSE across SNRs using the SDR baseline and the proposed SCA–SSD algorithm.}
}
\label{fig:comparison_K_and_L}
\end{figure}


\begin{figure}[!t]
\centering
\begin{tikzpicture}
    \begin{axis}[
        xlabel = {SNR},
        ylabel = {NMSE},
        label style={font=\footnotesize},
        width=0.42\textwidth,
        height=5.6cm,
        xmin=10, xmax=35,
        ymin=0.00025, ymax=0.1,
        legend style={nodes={scale=0.65, transform shape}, at={(0.3,0.85)}},
        ticklabel style = {font=\footnotesize},
        legend pos=south west,
        ymajorgrids=true,
        xmajorgrids=true,
        grid style=dashed,
        grid=both,
        ymode = log,
        grid style={line width=.1pt, draw=gray!10},
        major grid style={line width=.2pt,draw=gray!30},
    ]
    \addplot[ smooth,
             thin,
        color=chestnut,
        mark=square,
        line width=0.75pt,
        mark size=2pt,
        ]
    table[x=SNR,y=bs_prod]
    {Data/Sim_fading.dat};
    \addplot[ smooth,
             thin,
        color=airforceblue,
        mark=triangle,
        mark options = {rotate = 180, solid},
        line width=0.75pt,
        mark size=2.5pt,
        ]
    table[x=SNR,y=ReMAC_prod]
    {Data/Sim_fading.dat};
    \addplot[ smooth,
             thin,
        color=cssgreen,
        mark=triangle,
        line width=0.75pt,
        mark size=2.5pt,
        ]
    table[x=SNR,y=SeMAC_prod_SCA]
    {Data/Sim_fading.dat};
    \addplot[ smooth,
             thin,
        color=cadmiumorange,
        mark=+,
        mark options = {rotate = 45,solid},
        line width=0.75pt,
        mark size=2.5pt,
        ]
    table[x=SNR,y=channelComp_prod]
    {Data/Sim_fading.dat};
\legend{Bit-Slicing ,ReMAC ,SeMAC, ChannelComp};
\end{axis}
\end{tikzpicture}
\vspace{-1em}
\caption{Performance comparison among ReMAC, \ac{SeMAC}, ChannelComp and Bit-Slicing with \( K = 4 \) nodes for computing the product function.
}
\label{fig:comparison_L2}
\end{figure}


\vspace{-1em}
\subsection{Comparison to ChannelComp, ReMAC and Bit-Slicing}

In this subsection, we compare ChannelComp with repetition, \ac{SeMAC}, ReMAC, and Bit-Slicing for computing the product function with $K=4$ nodes over $L=2$ time slots, where input values are selected from $x_k \in \{1, 2, \ldots, 16\}$.
As shown in Figure~\ref{fig:comparison_L2}, by exploiting diverse modulation patterns across time slots, \ac{SeMAC} achieves lower \ac{NMSE} by approximately 14~dB compared to ReMAC. While ChannelComp with repetition and ReMAC offer similar performance, ReMAC achieves greater energy efficiency by selectively repeating modulated symbols. Moreover, Bit-Slicing yields the highest \ac{NMSE}, due to approximation errors introduced during the pre- and post-processing required to support nonlinear functions, as previously analyzed in~\cite{yan2025remac}.

\subsection{Performance of \ac{SeMAC-PA}}

We analyze the performance of \ac{SeMAC-PA} under Rayleigh fading with $K=8$ and \ac{QAM} modulation of orders $Q=4$ and $Q=16$ for computing the sum function over $L=2$ time slots.  
Figure~\ref{fig:power_adaptation} shows that \ac{NMSE} decreases with increasing \ac{SNR}, demonstrating the effectiveness of power adaptation. However, \ac{SeMAC-PA} relies on fixed QAM constellations, making it less effective than the optimized modulation design in \ac{SeMAC} under channel inversion power control.

\begin{figure}[!t]
\centering
\begin{tikzpicture}
    \begin{axis}[
        xlabel = {SNR},
        ylabel = {NMSE},
        label style={font=\footnotesize},
        width=0.42\textwidth,
        height=5.6cm,
        xmin=10, xmax=35,
        ymin=0.000003, ymax=0.85,
        legend style={nodes={scale=0.65, transform shape}, at={(0.3,0.85)}},
        ticklabel style = {font=\footnotesize},
        legend pos=south west,
        ymajorgrids=true,
        xmajorgrids=true,
        grid style=dashed,
        grid=both,
        ymode = log,
        grid style={line width=.1pt, draw=gray!10},
        major grid style={line width=.2pt,draw=gray!30},
    ]
    \addplot[smooth,
             thin,
             dashed,
        color=chestnut,
        mark=*,
        mark options = {rotate = 180, solid},
        line width=0.75pt,
        mark size=1.5pt,
        ]
    table[x=SNR,y=Q4L2]
    {Data/Sim_PA_SCA.dat};
    \addplot[smooth,
              thin,
            color=chestnut,
            mark=*,
            line width=0.75pt,
            mark size=1.5pt,
            ]
    table[x=SNR,y=Q4L2_PA]
    {Data/Sim_PA_SCA.dat};
    \addplot[smooth,
             thin, dashed,
        color=airforceblue,
        mark=square,
        mark options = {rotate = 45,solid},
        line width=0.75pt,
        mark size=2pt,
        ]
    table[x=SNR,y=Q16L2]
    {Data/Sim_PA_SCA.dat};
    \addplot[ smooth,
             thin,
        color=airforceblue,
        mark=square,
        mark options = {rotate = 45,solid},
        line width=0.75pt,
        mark size=2pt,
        ]
    table[x=SNR,y=Q16L2_PA]
    {Data/Sim_PA_SCA.dat};
\legend{SeMAC$\quad Q=4$,SeMAC-PA $Q=4$,SeMAC $Q=16$,SeMAC-PA $Q=16$};
\end{axis}
\end{tikzpicture}
\vspace{-1em}
\caption{Performance of \ac{SeMAC-PA} is evaluated across various SNRs in terms of \ac{NMSE} with \( K = 8 \) nodes.
}
\label{fig:power_adaptation}
\end{figure}


\section{Conclusion}\label{sec:conclusion}


In this letter, we proposed \ac{SeMAC}, a multi-symbol modulation framework for digital \ac{AirComp} that enables reliable function computation through distinct modulation patterns across multiple time slots. To support scenarios with fixed modulation, we introduced \ac{SeMAC-PA}, which adapts transmit power and phase to reshape the received constellation. {\color{black}One possible future work is to extend \ac{SeMAC} to scenarios with imperfect \ac{CSI} and investigate its integration with machine learning techniques for adaptive modulation design.}


\bibliographystyle{ieeetr}
\bibliography{Ref}

\begin{thebibliography}{10}

\bibitem{letaief2019roadmap}
K.~B. Letaief, W.~Chen, Y.~Shi, J.~Zhang, and Y.-J.~A. Zhang, ``The roadmap to 6{G}: {AI} empowered wireless networks,'' {\em IEEE Commun. Mag.}, vol.~57, no.~8, pp.~84--90, 2019.

\bibitem{wang2023road}
C.-X. Wang, X.~You, X.~Gao, X.~Zhu, Z.~Li, C.~Zhang, H.~Wang, Y.~Huang, Y.~Chen, H.~Haas, {\em et~al.}, ``On the road to 6{G}: Visions, requirements, key technologies and testbeds,'' {\em IEEE Commun. Surveys Tuts.}, 2023.

\bibitem{csahin2023survey}
A.~{\c{S}}ahin and R.~Yang, ``A survey on over-the-air computation,'' {\em IEEE Commun. Surveys Tuts.}, 2023.

\bibitem{zhu2018mimo}
G.~Zhu and K.~Huang, ``{MIMO} over-the-air computation for high-mobility multimodal sensing,'' {\em IEEE Internet Things J.}, vol.~6, no.~4, pp.~6089--6103, 2018.

\bibitem{saeed2023ChannelComp}
S.~Razavikia, J.~M.~B. Da~Silva~Jr, and C.~Fischione, ``Channel{Comp}: A general method for computation by communications,'' {\em IEEE Trans. Commun.}, vol.~72, no.~2, pp.~692--706, 2024.

\bibitem{yan2025remac}
X.~Yan, S.~Razavikia, and C.~Fischione, ``Re{MAC}: Digital multiple access computing by repeated transmissions,'' {\em IEEE Trans. Commun.}, vol.~73, no.~10, pp.~8965--8979, 2025.

\bibitem{yan2024novel}
X.~Yan, S.~Razavikia, and C.~Fischione, ``A novel channel coding scheme for digital multiple access computing,'' in {\em IEEE International Conference on Communications}, pp.~3851--3857, 2024.

\bibitem{liu2025digital}
J.~Liu, Y.~Gong, and K.~Huang, ``Digital over-the-air computation: Achieving high reliability via bit-slicing,'' {\em IEEE Trans. Wireless Commun.}, vol.~24, no.~5, pp.~4101--4114, 2025.

\bibitem{luo2010semidefinite}
Z.-Q. Luo, W.-K. Ma, A.~M.-C. So, Y.~Ye, and S.~Zhang, ``Semidefinite relaxation of quadratic optimization problems,'' {\em IEEE Signal Process. Mag.}, vol.~27, no.~3, pp.~20--34, 2010.

\bibitem{guo2021over}
H.~Guo, Y.~Zhu, H.~Ma, V.~K. Lau, K.~Huang, X.~Li, H.~Nong, and M.~Zhou, ``Over-the-air aggregation for federated learning: Waveform superposition and prototype validation,'' {\em J. Commun. Inf. Netw.}, vol.~6, no.~4, pp.~429--442, 2021.

\bibitem{cao2019optimal}
X.~Cao, G.~Zhu, J.~Xu, and K.~Huang, ``Optimal power control for over-the-air computation,'' in {\em IEEE Global Communications Conference}, pp.~1--6, 2019.

\bibitem{beck2010sequential}
A.~Beck, A.~Ben-Tal, and L.~Tetruashvili, ``A sequential parametric convex approximation method with applications to nonconvex truss topology design problems,'' {\em J. of Global Optim.}, vol.~47, no.~1, pp.~29--51, 2010.

\bibitem{konar2017first}
A.~Konar and N.~D. Sidiropoulos, ``First-order methods for fast feasibility pursuit of non-convex {QCQP}s,'' {\em IEEE Trans. Signal Process.}, vol.~65, no.~22, pp.~5927--5941, 2017.

\end{thebibliography}



\end{document}